\documentclass[aip,jcp,reprint]{revtex4-1}
\usepackage{graphicx}
\usepackage{dcolumn}
\usepackage{color}
\usepackage{amsfonts,amsmath,latexsym,bbm,dsfont}

\usepackage{etoolbox}
\usepackage{multirow}

\makeatletter
\def\@email#1#2{%
 \endgroup
 \patchcmd{\titleblock@produce}
  {\frontmatter@RRAPformat}
  {\frontmatter@RRAPformat{\produce@RRAP{*#1\href{mailto:#2}{#2}}}\frontmatter@RRAPformat}
  {}{}
}%
\makeatother

\newcommand{\bra}[1]{\langle{#1}|}
\newcommand{\ket}[1]{|{#1}\rangle}
\newcommand{\TwoEl}[2]{\langle{#1}{|}{#2}\rangle}

\newcommand{\TwoElChem}[2]{({#1}{|}{#2})}

\renewcommand{\vec}[1]{{\mathbf{#1}}}
\newcommand{\chg}[1]{{\color{blue} #1}}
\renewcommand{\chg}[1]{{#1}}

\begin{document}


\title{\chg{xTC:} An efficient treatment of three-body interactions in transcorrelated methods} 



\author{Evelin Martine Christlmaier}
\email[]{e.christlmaier@fkf.mpg.de}
\affiliation{Max Planck Institute for Solid State Research, Heisenbergstraße 1, 70569 Stuttgart, Germany}
\author{Thomas Schraivogel}
\affiliation{Max Planck Institute for Solid State Research, Heisenbergstraße 1, 70569 Stuttgart, Germany}
\author{Pablo L\'opez R\'ios}
\affiliation{Max Planck Institute for Solid State Research, Heisenbergstraße 1, 70569 Stuttgart, Germany}
\author{Ali Alavi}
\affiliation{Max Planck Institute for Solid State Research, Heisenbergstraße 1, 70569 Stuttgart, Germany}
\affiliation{Yusuf Hamied Department of Chemistry, University of Cambridge, Lensfield Road, Cambridge CB2 1EW, United Kingdom}
\author{Daniel Kats}
\email[]{d.kats@fkf.mpg.de}
\affiliation{Max Planck Institute for Solid State Research, Heisenbergstraße 1, 70569 Stuttgart, Germany}


\date{\today}

\begin{abstract}
An efficient implementation for approximate inclusion of the three-body operator arising in transcorrelated methods 
via exclusion of \chg{explicit three-body components} (xTC) 
is presented and tested against results in the ``HEAT'' benchmark set [A.\ Tajti \textit{et al.}, J.\ Chem.\ Phys.\ \textbf{121}, 11599 (2004)].
Using relatively modest basis sets and computationally simple methods, 
total, atomization, and formation energies within near-chemical accuracy from HEAT results were obtained.
The xTC ansatz reduces the nominal scaling of the three-body part of transcorrelation by two orders of magnitude to $\mathcal O(N^5)$
and can readily be used with almost any quantum chemical correlation method.
\end{abstract}

\maketitle 

\section{Introduction}
Achieving chemical accuracy in computational quantum chemistry calculations is still a significant challenge for all but the most simple systems.
Convergence of the correlation energy with basis set size is often slow
and high-quality correlation methods are required to obtain accurate results,
leading to rapidly escalating computational cost.

One of the reasons for the slow convergence of the energy with the size of the basis set
are the discontinuities in the first derivative of the wave function at the electron coalescence points $\vec r_{ij} = 0$,
\textit{i.e.}, the Kato cusp.
\cite{k1957}
Explicitly correlated methods aim to address this issue
by introducing explicit dependence on the inter-electronic distances into the wave function,
leading to significantly improved convergence to the basis set limit.
\cite{k1985,kk1991i,ks2002,m2003,tenno2004,tenno2004a,v2004,tk2005,kedzuch:05,fkh05,fhk06b,f12g,noga:07,ccf12,
tknh07,rmp2f12,skhv08a,skhv08b,nkst08,tkh2008,valeev08,valeev08b,Torheyden:2008,bokhan2008,kaw2009,Werner:2011,
Ten-no:TCA131-1,Hattig:CR112-4,Kong:CR112-75} 
However, extension to higher-order methods is non-trivial.

Alternatively, the electronic wavefunction $\Psi$ can be factorized into a Jastrow factor $e^\tau$ and a smoother function $\Phi$
\begin{eqnarray}
\Psi = e^\tau \Phi
\end{eqnarray}
with the correlation factor
\begin{eqnarray}
    \tau = \sum_{i<j} u(\vec r_i,\vec r_j)
\end{eqnarray}
describing the correlation of the electron pairs via the symmetric functions $u(\vec r_i, \vec r_j)$.
\cite{Jastrow}
Transferring the Jastrow factor to the Hamiltonian via a similarity transformation
\begin{eqnarray}
    e^{-\tau} H e^\tau\Phi = \chg{\tilde H} \Phi = E\Phi
\end{eqnarray}
gives rise to the \chg{non-hermitian transcorrelated Hamiltonian, $\tilde H$}. The transcorrelated ansatz was pioneered by Boys and Handy\cite{boysCalculation1969}
and has seen renewed interest in recent years.
\cite{ten-noFeasible2000,hinoApplication2002,umezawa_transcorrelated_2003,
yanai_canonical_2006,yanai_canonical_2007,yanai_canonical_2012,
tsuneyukiTranscorrelated2008,
ochi_efficient_2012,ochi_optical_2014,ochi_second-order_2015,ochi_iterative_2016,wahlen-strothman_lie_2015,luoCombining2018,
dobrautzCompact2019,cohenSimilarity2019,baiardi_transcorrelated_2020,khamoshi_exploring_2021,giner_new_2021,gutherberyliumdimer2021,liaoEfficient2021,TCCC2021,liao22,haupt23,TCCC2022} 
Compared to the explicitly correlated methods, the transcorrelated ansatz has the advantage
that it not only improves the basis set convergence, but also the quality of the underlying correlation method.

A major bottleneck of transcorrelated methods is the calculation of the three-body integrals arising from the
similarity transformation of the Hamiltonian. Further, they require the implementation of additional modifications
in the subsequent correlation methods that cost both additional CPU and developer time.
Recently, \chg{approximate transcorrelated} coupled cluster
\cite{cizek:66,Purvis:82,Raghavachari:89} 
and distinguishable cluster
\cite{kats_dc_2013,kats_dcsd_2014,kats_accurate_2015,kats_distinguishable_2016,kats_improving_2018}
methods that drop the pure three-body part of the
normal ordered 
\chg{similarity transformed}
Hamiltonian have been proposed and shown to introduce only negligible errors in the final energies
\chg{for atoms and molecules\cite{TCCC2021,TCCC2022} as well as the three-dimensional uniform electron gas\cite{liaoEfficient2021}.}
However, explicit calculation of the costly three-body integrals via numerical integration was still required.

In this work, we introduce an efficient algorithm for the inclusion of the transcorrelated
three-body operator excluding explicit \chg{three-body components} (xTC)
via modifications of the zero, one- and two-body integrals.
Instead of calculating the three-body integrals explicitly, 
the corrections to the integrals are obtained directly from the intermediates 
of the numerical integration, leading to a reduction of the scaling of the matrix-element evaluation stage by two orders of magnitude with respect to a na{\"\i}ve implementation.
Extensive benchmarking results using different correlation methods,  basis sets and additional approximations are presented.

The paper is organized as follows. In section~\ref{sec:theory} we will derive the working equations for the xTC ansatz in spin-orbital basis.
Section~\ref{sec:results} shows the results of the benchmarking calculations. Section~\ref{sec:conclusion} will offer a brief summary of the results.
Finally, spin-integrated working equations are shown in the appendix.

\section{\label{sec:theory}Theory}
\subsection{Transcorrelation}
The transcorrelated Hamiltonian in second quantization is
\begin{align}
    \chg{\tilde H} =& h_P^Q a^P_Q
           + \frac{1}{2}(V_{PR}^{QS}-K_{PR}^{QS}) a^{PR}_{QS}
           \nonumber\\&
           - \frac{1}{6}L_{PRT}^{QSU} a^{PRT}_{QSU}
           + E_\text{nuc}
\end{align}
where $P$, $Q$, $R$, ... are general spin-orbital indices
and 
\begin{subequations}
\begin{eqnarray}
    a^P_Q = a^P a_Q = a_P^\dagger a_Q
\end{eqnarray}
\begin{eqnarray}
    a^{PR}_{QS} = a^P a^R a_S a_Q
\end{eqnarray}
\begin{eqnarray}
    a^{PRT}_{QSU} = a^P a^R a^T a_U a_S a_Q.
\end{eqnarray}
\end{subequations}
are products of creation and annihilation operators.
Note that here and throughout the article we use the Einstein convention, \textit{i.e.}, we implicitly sum over repeated indices
on the right hand side of the equation as long as they do not appear on the left hand side.

In addition to the standard integrals
\begin{subequations}
\begin{eqnarray}
    \label{eq:Inth}
    h_P^Q = \bra{P}h\ket{Q}
\end{eqnarray}
\begin{eqnarray}
    \label{eq:IntV}
    V_{PR}^{QS} = \TwoEl{PR}{QS} = \TwoElChem{PQ}{RS}
    ,
\end{eqnarray}
\end{subequations}
the \chg{Baker-Campbell-Hausdorff} expansion of the similarity transformed Hamiltonian
gives rise to the non-hermitian two-body integrals
\begin{align}
  K_{PR}^{QS} = \frac{1}{2}\bra{PR}
    \mathcal P^1_2(&
  \nabla_1^2 u(\vec r_1,\vec r_2)
  +(\nabla_1 u(\vec r_1,\vec r_2))^2
    \nonumber\\&
  +2\nabla_1 u(\vec r_1,\vec r_2)\cdot\nabla_1
  )
  \ket{QS}
\end{align}
and the three-body integrals
\begin{align}
    L_{PRT}^{QSU}=\bra{PRT}\mathcal P_{2,3}^1 (&\nabla_1 u(\vec r_1,\vec r_2)
    \nonumber\\&\cdot \nabla_1 u(\vec r_1,\vec r_3))\ket{QSU}
\end{align}
where we have used the symmetric permutation operators
\begin{subequations}
\begin{eqnarray}
    \mathcal P^1_2 f(1,2) = f(1,2) + f(2,1)
\end{eqnarray}
\begin{align}
    \mathcal P^1_{2,3} f(1,2,3) = f&(1,2,3) + f(2,1,3)
    \nonumber\\&+ f(3,2,1)
    .
\end{align}
\end{subequations}

\chg{Aside from evaluation of the costly three-body integrals, 
inclusion of the three-body operator $L_N$ requires modification
of any theory using transcorrelation to include terms interacting
with a three-body operator in the Hamiltonian.
In order to find approximations that will solve
these issues, we will first introduce normal ordering 
in the next section.}

\subsection{Generalized normal ordering}
Instead of defining normal-order with respect to a single determinant,
Kutzelnigg and Mukherjee proposed a generalized normal ordering with respect
to an arbitrary reference function $\ket{\Phi_0}$.
\cite{KutzMuk1997,KutzMuk1999}

In the general normal ordering, the antisymmetrized products of Kronecker deltas are replaced by the density matrices 
of the reference function
\begin{eqnarray}
    \gamma^{P...}_{Q...} = \bra{\Phi_0}a^{P...}_{Q...}\ket{\Phi_0}.
\end{eqnarray}
For a three-body operator, this yields
\begin{align}
    a^{PRT}_{QSU} =& \tilde a^{PRT}_{QSU}
    + \sum (-1)^{\mathcal P} \gamma^{P}_{Q}\tilde a^{RT}_{SU}
    \nonumber\\&
    + \sum (-1)^{\mathcal P} \gamma^{PR}_{QS}\tilde a^{T}_{U}
    + \gamma^{PRT}_{QSU}
\end{align}
with $\tilde a$ being normal ordered operators and the sums 
going over all nine non-redundant permutations of (PRT) and (QSU) with the appropriate sign.

Introducing the combined two-electron integrals
\begin{eqnarray}
    U_{PR}^{QS} = V_{PR}^{QS} - K_{PR}^{QS}
\end{eqnarray}
the normal ordered Hamiltonian can be separated into 
effective one-, two- and three-body operators
\begin{eqnarray}
    \chg{\tilde H_N = \tilde H - \bra{\Phi_0} \tilde H\ket{\Phi_0} =} F_N + V_N + L_N
\end{eqnarray}
with
\begin{subequations}
\begin{align}
    F_N =
    \bigg[&
        h_P^Q
        +\big(U_{PR}^{QS}-U_{PR}^{SQ}\big)\gamma^R_S
        \nonumber\\&
        -\frac{1}{2}\big(L_{PRT}^{QSU}-L_{PRT}^{SQU}-L_{PRT}^{USQ}\big)\gamma^{RT}_{SU}
    \bigg]\tilde a^{P}_{Q}
\end{align}
\begin{align}
    V_N =\frac{1}{2} \bigg[&
        U_{PR}^{QS}
        -\big(L_{PRT}^{QSU}
        \nonumber\\&
        -L_{PRT}^{QUS}-L_{PRT}^{USQ}\big)\gamma^T_U
    \bigg]\tilde a^{PR}_{QS}
\end{align}
\begin{eqnarray}
    \label{eq:LN}
    L_N =-\frac{1}{6} L_{PRT}^{QSU}\tilde a^{PRT}_{QSU}
\end{eqnarray}
\end{subequations}

\subsection{The xTC approximation: excluding \chg{explicit three-body components}}
\chg{We can now exclude the explicit three-body components}
$L_N$ (xTC) and incorporate the remaining 3-body contributions 
via a change of the two-, one- and zero-body integrals.
\begin{subequations}
\label{eq:xTC}
\begin{eqnarray}
    \bar U_{PR}^{QS}\leftarrow U_{PR}^{QS}+\Delta U_{PR}^{QS} 
\end{eqnarray}
\begin{eqnarray}
    \bar h_{P}^{Q}\leftarrow h_{P}^{Q}+\Delta h_{P}^{Q}        
\end{eqnarray}
\begin{eqnarray}
    \bar E_\text{nuc} \leftarrow E_\text{nuc} + \bra{\Phi_0}L\ket{\Phi_0}
\end{eqnarray}
\end{subequations}
Neglect of the pure normal ordered three-body operator eq. (\ref{eq:LN}) has been shown to introduce only minor errors,
even for normal ordering with respect to the Hartree-Fock determinant.
\cite{TCCC2021,TCCC2022}

The correction to the two-electron integrals is trivially obtained as
\begin{eqnarray}
    \label{eq:dU}
    \Delta U_{PR}^{QS} = -\big(L_{PRT}^{QSU}-L_{PRT}^{QUS}-L_{PRT}^{USQ}\big)\gamma^T_U
    .
\end{eqnarray}
Inserting $\bar U$ instead of $U$ into $F_N$ yields a new generalized Fock-operator
\begin{align}
    F_N^\prime =
    \bigg[
        h_P^Q 
        +&\big(U_{PR}^{QS}-U_{PR}^{SQ}\big)\gamma^R_S
        \nonumber\\ 
        +&\big(L_{PRT}^{QSU}-L_{PRT}^{SQU}-L_{PRT}^{USQ}\big)
        \nonumber\\&\times
        \big(-\gamma^R_S\gamma^T_U+\gamma^R_U\gamma^T_S-\frac{1}{2}\gamma^{RT}_{SU}\big)
    \bigg]\tilde a^{P}_{Q}
\end{align}
that differs from the original generalized Fock-operator $F_N$ according to
\begin{align}
    F_N = F_N^\prime 
    -& \big(L_{PRT}^{QSU}-L_{PRT}^{SQU}-L_{PRT}^{USQ}\big)
    \nonumber\\&\times
    \big(-\gamma^R_S\gamma^T_U+\gamma^R_U\gamma^T_S\big))\tilde a^{P}_{Q}
    .
\end{align}

In order to ensure that the generalized Fock-operator remains invariant with respect to
the change in integrals, we combine the correction of the change
due to $\Delta U_{PR}^{QS}$ in $F_N$ 
with the one-body correction arising due to the normal ordering of the three-body operator
to obtain
\begin{align}
    \Delta h_P^Q = 
        &\big(L_{PRT}^{QSU}-L_{PRT}^{SQU}-L_{PRT}^{USQ}\big)
        \nonumber\\&\times
        \big(\gamma^R_S\gamma^T_U-\gamma^R_U\gamma^T_S-\frac{1}{2}\gamma^{RT}_{SU}\big)
        .
\end{align}

The zero-body correction is simply the expectation value of the three-body operator
\begin{eqnarray}
 \bra{\Phi_0}L\ket{\Phi_0} = - \frac{1}{6}L_{PRT}^{QSU}\gamma^{PRT}_{QSU}
    .
\end{eqnarray}

\chg{So far these equations are valid for an arbitrary reference function $\ket{\Phi_0}$. 
In the next section we will show how a single-determinant reference
function greatly simplifies the equations.}

\subsection{Single determinant case}
If the reference function $\ket{\Phi_0}$ is a single determinant, 
the higher order density matrices are simply antisymmetrized products of the one-body density matrix
\begin{eqnarray}
    \label{eq:1Ddens}
    \gamma^{PR...}_{QS...} = \mathcal A (\gamma^P_Q\gamma^R_S...).
\end{eqnarray}

In this case, the one- and zero-body corrections are trivially obtained from 
contraction of the higher-body correction with the one-body density matrix
\begin{align}
    \label{eq:tcmf:1el}
    \Delta h_P^Q =&
        \frac{1}{2}\big(L_{PRT}^{QSU}-L_{PRT}^{SQU}-L_{PRT}^{USQ}\big)
        \nonumber\\&\times\big(\gamma^R_S\gamma^T_U-\gamma^R_U\gamma^T_S\big)
        \nonumber\\=&
        -\frac{1}{2}\big(\Delta U_{PR}^{QS}-\Delta U_{PR}^{SQ}\big)\gamma^{R}_{S}
\end{align}

\begin{eqnarray}
    \label{eq:tcmf:0el}
    \bra{\Phi_0}L\ket{\Phi_0}=-\frac{1}{3}\Delta h_{P}^{Q}\gamma^{P}_{Q}
\end{eqnarray}

While these two equations are only exact for a single-determinant reference, 
they can still be used approximately with correlated densities (\textit{i.e.}, negelectng the higher order cumulants in the cumulant expansion of the respective density\cite{KutzMuk1997,KutzMuk1999}),  
which offers the potential for additional flexibility in the xTC ansatz.

\chg{The equations given above still require explicit calculation of the L-matrix. 
In the next section we will describe how we can obtain the corrections in (\ref{eq:xTC})
without the need of costly six-index intermediates.}

\subsection{Decomposition of two-body correction}
In our implementation\cite{cohenSimilarity2019}, the three-body integrals are obtained via numerical integration
over the grid points $x_i$ with the weights $w_i$
\begin{align}
    \label{eq:Ldecomp}
    L_{PRT}^{QSU}=
      P^{(PQ)}_{(RS),(TU)}
      \sum_i w_i &\phi_P^*(i)
      \vec V_{R}^{S}(i)
      \cdot\vec V_{T}^{U}(i)\phi^Q(i)
\end{align}
and the intermediate
\begin{eqnarray}
    \vec V_{R}^{S}(i) = \sum_j w_j \phi_R^*(j)\big(\nabla u(\vec r_i,\vec r_j)\big) \phi^S(j)
    .
\end{eqnarray}
We use boldface here to indicate that $\vec V_{R}^{S}(i)$ and some of the following intermediates are three-dimensional spatial vectors 
and use the scalar product to indicate contraction over the spatial coordinates.

Making use of the decomposition eq. (\ref{eq:Ldecomp}) allows us to avoid calculation of the three-body integrals entirely
and instead calculate $\Delta U$ directly.
By inserting eq. (\ref{eq:1Ddens}) into eq. (\ref{eq:dU})
and changing the order of summation, we obtain
\begin{align}
    \label{eq:tcmf:2el}
    \Delta U_{PR}^{QS} =
        -\mathcal P^{(PQ)}_{(RS)}
        \big(&
        \rho_{P}^{Q}(i)A_{R}^{S}(i)
        +
        \vec V_{P}^{Q}(i)\cdot\vec B_{R}^{S}(i)
        \big)
\end{align}
with the intermediates
\begin{subequations}
    \label{eq:tcmf:inter}
\begin{eqnarray}
    \rho_{P}^{Q}(i) =w_i\phi_P^*(i)\phi^Q(i)
\end{eqnarray}
\begin{eqnarray}
    A_{R}^{S}(i) = \tilde V_{R}^{S}(i) - \tilde Z_{R}^{S}(i)
\end{eqnarray}
\begin{eqnarray}
    \tilde V_{R}^{S}(i) = \vec W(i)\cdot\vec V_{R}^{S}(i)
\end{eqnarray}
\begin{eqnarray}
    \vec W(i) = \vec V_{T}^{U}(i)\gamma^T_{U}
\end{eqnarray}
\begin{eqnarray}
    \tilde Z_{R}^{S}(i) = \vec V_{R}^{U}(i)\cdot\vec X^{S}_{U}(i)
\end{eqnarray}
\begin{eqnarray}
    \vec X^{S}_{U}(i) = \vec V_{T}^{S}(i) \gamma^T_U
\end{eqnarray}
\begin{eqnarray}
    \vec B_{R}^{S}(i) = \frac{1}{2}\tilde W(i)\vec V_{R}^{S}(i) - \vec Z_{R}^{S}(i)
\end{eqnarray}
\begin{eqnarray}
    \tilde W(i) = \rho_{T}^{U}(i)\gamma^T_U
\end{eqnarray}
\begin{eqnarray}
    \vec Z_{R}^{S}(i) = \rho_{R}^{U}(i)\vec X^{S}_{U}(i) + \vec Y_{R}^{T}(i) \rho_{T}^{S}(i)
\end{eqnarray}
\begin{eqnarray}
    \vec Y_{R}^{T}(i) = \vec V_{R}^{U}(i) \gamma^T_U
\end{eqnarray}
\end{subequations}

Whereas the formal scaling of evaluation of the three-body integrals is of order $N_\text{orb}^6 N_\text{grid}$,
the most expensive step of the xTC ansatz presented here, \textit{i.e.}, the assembly of the two-body correction
eq. (\ref{eq:tcmf:2el}), scales only as $N_\text{orb}^4 N_\text{grid}$. 

\section{\label{sec:results}Results}
\subsection{Computational details}
The ``HEAT'' series of papers \cite{HEAT1, HEAT2, HEAT3, HEAT4} provide
bechmark-quality energetics for 31 atoms and molecules obtained with
computationally expensive, accurate correlation methods and large
basis sets, employing basis-set extrapolation to produce the final
results.
We compute the total, atomization, and formation energies of these
systems using the transcorrelation ansatz excluding explicit \chg{three-body components} (xTC) with
moderate basis-set sizes without extrapolation and compare them
against the corresponding nonrelativistic electronic energies from
HEAT, reported as Hartree-Fock and correlation energy contributions in
Table I of Ref. \onlinecite{HEAT1}, which we treat as ``exact''
values.

Note that carbon is a solid under standard conditions, so ``proper''
formation energies of compounds containing carbon atoms cannot be
calculated from our results.
For the purpose of comparing with Table~III of Ref.\
\onlinecite{HEAT1} we compute formation energies with respect to
carbon monoxide instead, \textit{e.g.}, the formation energy for
C$_2$H$_2$ is obtained from the reaction
\begin{eqnarray}
  2\text{CO}+\text{H$_2$}\rightarrow\text{C$_2$H$_2$} + 2 \text{O} \;.
\end{eqnarray}
Table~\ref{tab:form} lists all reactions considered in the evaluation
of formation energies.

\begin{table}
\caption{\label{tab:form} Chemical reactions used for formation energies.}
\begin{ruledtabular}
\begin{tabular}{ccc}
{CO - O                                                       }&{$\rightarrow$}&{C          }\\      
{$\frac{1}{2}$F$_2$                                           }&{$\rightarrow$}&{F          }\\
{$\frac{1}{2}$H$_2$                                           }&{$\rightarrow$}&{H          }\\
{$\frac{1}{2}$N$_2$                                           }&{$\rightarrow$}&{N          }\\
{$\frac{1}{2}$O$_2$                                           }&{$\rightarrow$}&{O          }\\
{$2$CO - $2$O + H$_2$                                         }&{$\rightarrow$}&{C$_2$H$_2$ }\\
{$2$CO - $2$O + $\frac{1}{2}$H$_2$                            }&{$\rightarrow$}&{CCH        }\\
{CO - O + H$_2$                                               }&{$\rightarrow$}&{CH$_2$     }\\
{CO - O + $\frac{1}{2}$H$_2$                                  }&{$\rightarrow$}&{CH         }\\
{CO - O + $\frac{3}{2}$H$_2$                                  }&{$\rightarrow$}&{CH$_3$     }\\
{CO - O + O$_2$                                               }&{$\rightarrow$}&{CO$_2$     }\\
{H$_2$ + O$_2$                                                }&{$\rightarrow$}&{H$_2$O$_2$ }\\
{H$_2$ + $\frac{1}{2}$O$_2$                                   }&{$\rightarrow$}&{H$_2$O     }\\
{CO - O + $\frac{1}{2}$H$_2$ + $\frac{1}{2}$O$_2$             }&{$\rightarrow$}&{HCO        }\\
{$\frac{1}{2}$H$_2$ + $\frac{1}{2}$F$_2$                      }&{$\rightarrow$}&{HF         }\\
{$\frac{1}{2}$H$_2$ + O$_2$                                   }&{$\rightarrow$}&{HO$_2$     }\\
{$\frac{1}{2}$N$_2$ + $\frac{1}{2}$O$_2$                      }&{$\rightarrow$}&{NO         }\\
{$\frac{1}{2}$H$_2$ + $\frac{1}{2}$O$_2$                      }&{$\rightarrow$}&{OH         }\\
{$\frac{1}{2}$H$_2$ + $\frac{1}{2}$N$_2$ + $\frac{1}{2}$O$_2$ }&{$\rightarrow$}&{HNO        }\\
{CO - O + $\frac{1}{2}$N$_2$                                  }&{$\rightarrow$}&{CN         }\\
{$\frac{1}{2}$H$_2$ + CO - O + $\frac{1}{2}$N$_2$             }&{$\rightarrow$}&{HCN        }\\
{CO - O + $\frac{1}{2}$F$_2$                                  }&{$\rightarrow$}&{CF         }\\
{$\frac{1}{2}$N$_2$ + H$_2$                                   }&{$\rightarrow$}&{NH$_2$     }\\
{$\frac{1}{2}$N$_2$ + $\frac{3}{2}$H$_2$                      }&{$\rightarrow$}&{NH$_3$     }\\
{$\frac{1}{2}$N$_2$ + $\frac{1}{2}$H$_2$                      }&{$\rightarrow$}&{NH         }\\
{$\frac{1}{2}$O$_2$ + $\frac{1}{2}$F$_2$                      }&{$\rightarrow$}&{OF         }
\end{tabular}
\end{ruledtabular}
\end{table}

For the Jastrow factors, we employ the Drummond-Towler-Needs form 
\chg{
\begin{align}
    u(\vec r_i,\vec r_j) =& 
    v(r_{ij})
    + \frac{1}{N_{el}-1}
    \sum_I \big[
        \chi(r_{iI})
        +
        \chi(r_{jI})
    \big]
    \nonumber\\&
    + f(r_{ij}, r_{iI}, r_{jI})
\end{align}
}
with electron-electron,
electron-nucleus, and electron-electron-nucleus contributions
$v$, $\chi$, and $f$, respectively, expanded in natural powers.
\cite{drummond_jastrow_2004, lopezrios_jastrow_2012}
The Jastrow functions have been optimized by minimizing the variance
of the reference energy within variational Monte Carlo,
with the details being described elsewhere.
\cite{haupt23}

Numerical quadrature over the direct product of atom centered grids built from 
Treutler-Ahlrichs radial grids and Lebedev angular grids obtained from PySCF has been employed.
\cite{PySCF}
We observe that results are essentially converged for grid level 2 (see supplementary material) and this is the choice we have used throughout the paper.
Final results were obtained using
unrestriced coupled cluster (CC) and distinguishable cluster (DC) methods implemented
in the Molpro quantum chemistry package.
\cite{MOLPRO}

In the following, we investigate the quality of the energies obtained 
with different coupled cluster based correlation methods as well 
as different types and sizes of basis sets.
A comparison to the well established explicitly correlated methods represented by the F12a approximation
\chg{with the diagonal ansatz 3C(D) as implemented in Molpro\cite{knizia09,kats15,kedzuch05,werner07}}
has been included as well.
Finally, we investigate the effect of further approximations to the xTC ansatz itself.
Unless otherwise specified, the density matrix used in the xTC contractions is the HF density 
as described in appendix~\ref{sec:ROHFwork} for open-shell molecules and appendix~\ref{sec:RHFwork} for closed-shell molecules,
which corresponds to approximation~B in previous work.\cite{TCCC2021,TCCC2022}
Furthermote, if not otherwise indicated, all-electron calculations have been carried out.

In order to simplify the comparison, we will only present 
the signed mean error (ME), standard deviation (STD) and maximum error (MaxE) for total energies
and the mean absolute error (MAE), root mean square error (RMSE) and MaxE for relative energies in 
the body of this work. Individual results and additional figures can be found in the supplementary material.


\subsection{Method dependence}
\begin{figure*}
\includegraphics[width=.95\textwidth,keepaspectratio]{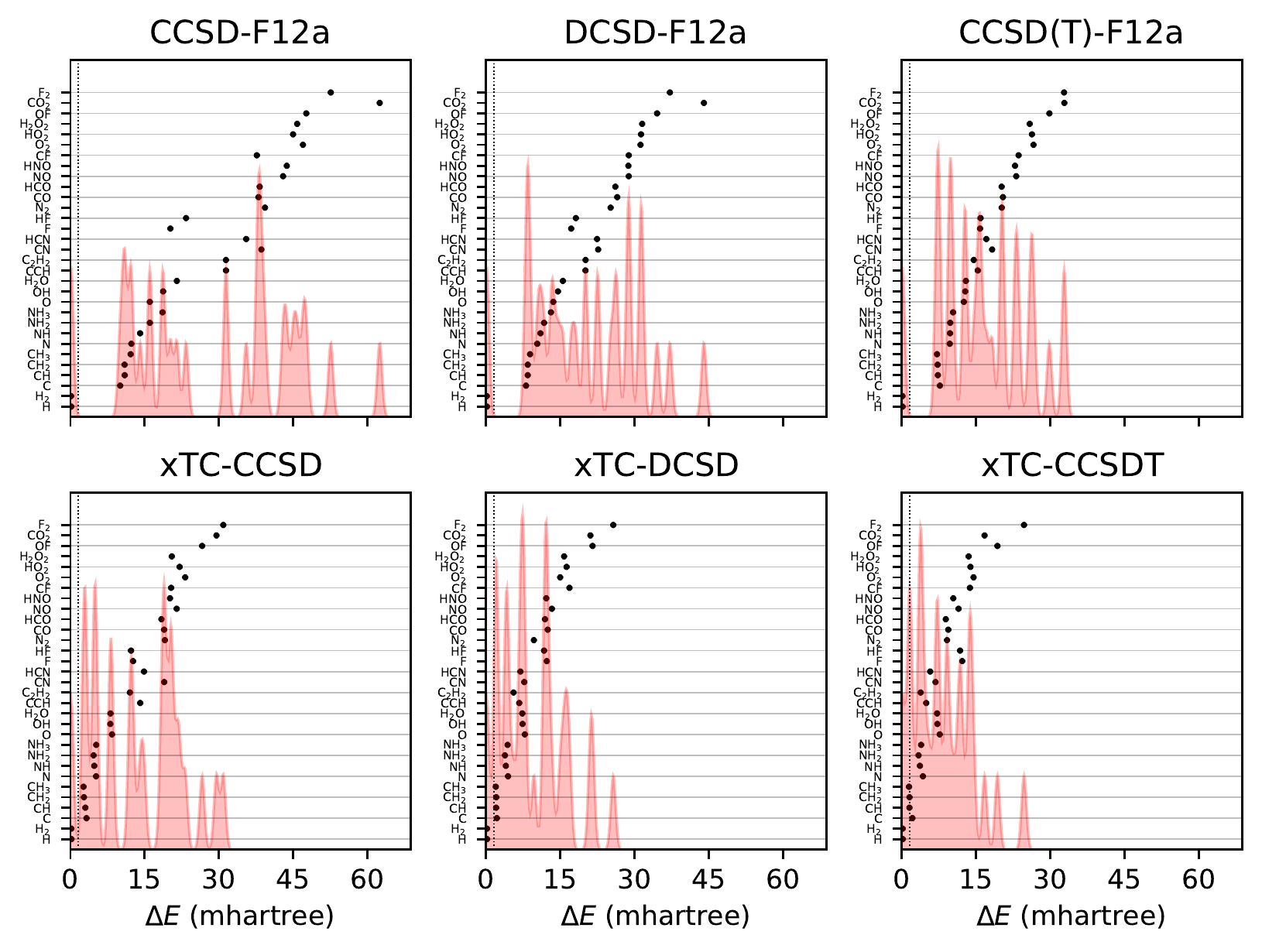}
\caption{\label{plt:methods:mol} Method dependence of total energies (aug-cc-pVTZ) compared to HEAT.
    Dotted lines indicate chemical accuracy.
    The shaded area corresponds to the sum of gaussians centered at the data points
    with the width of the gaussians chosen such that equidistantly distributed gaussians would be \chg{contained} to 95\% in the corresponding segment.
}
\end{figure*}
\begin{figure*}
\includegraphics[width=.95\textwidth,keepaspectratio]{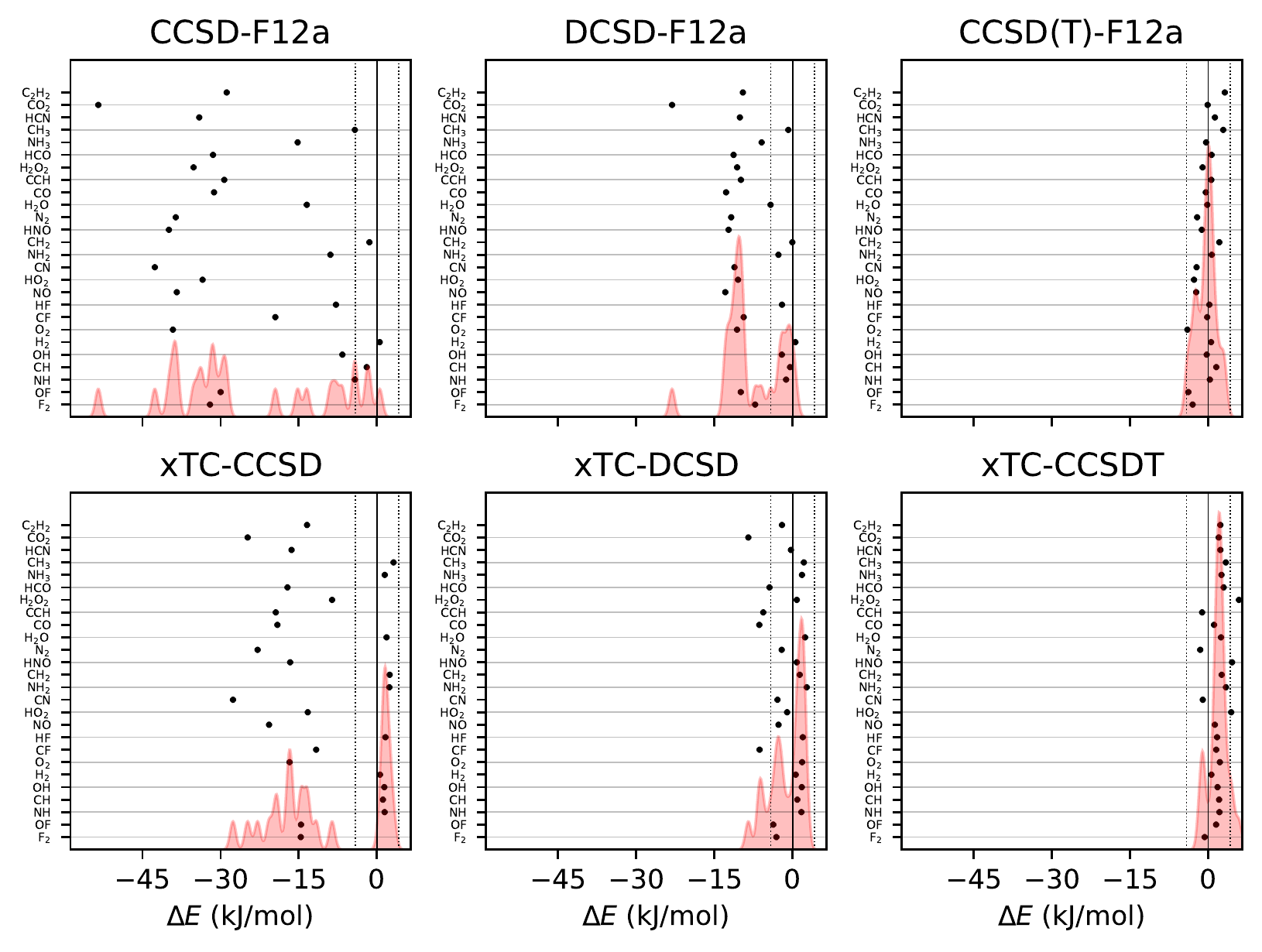}
\caption{\label{plt:methods:at} Method dependence of atomization energies (aug-cc-pVTZ) compared to HEAT.
    Dotted lines indicate chemical accuracy.
    The shaded area corresponds to the sum of gaussians centered at the data points
    with the width of the gaussians chosen such that equidistantly distributed gaussians would be \chg{contained} to 95\% in the corresponding segment.
    }
\end{figure*}
\begin{figure*}
\includegraphics[width=.95\textwidth,keepaspectratio]{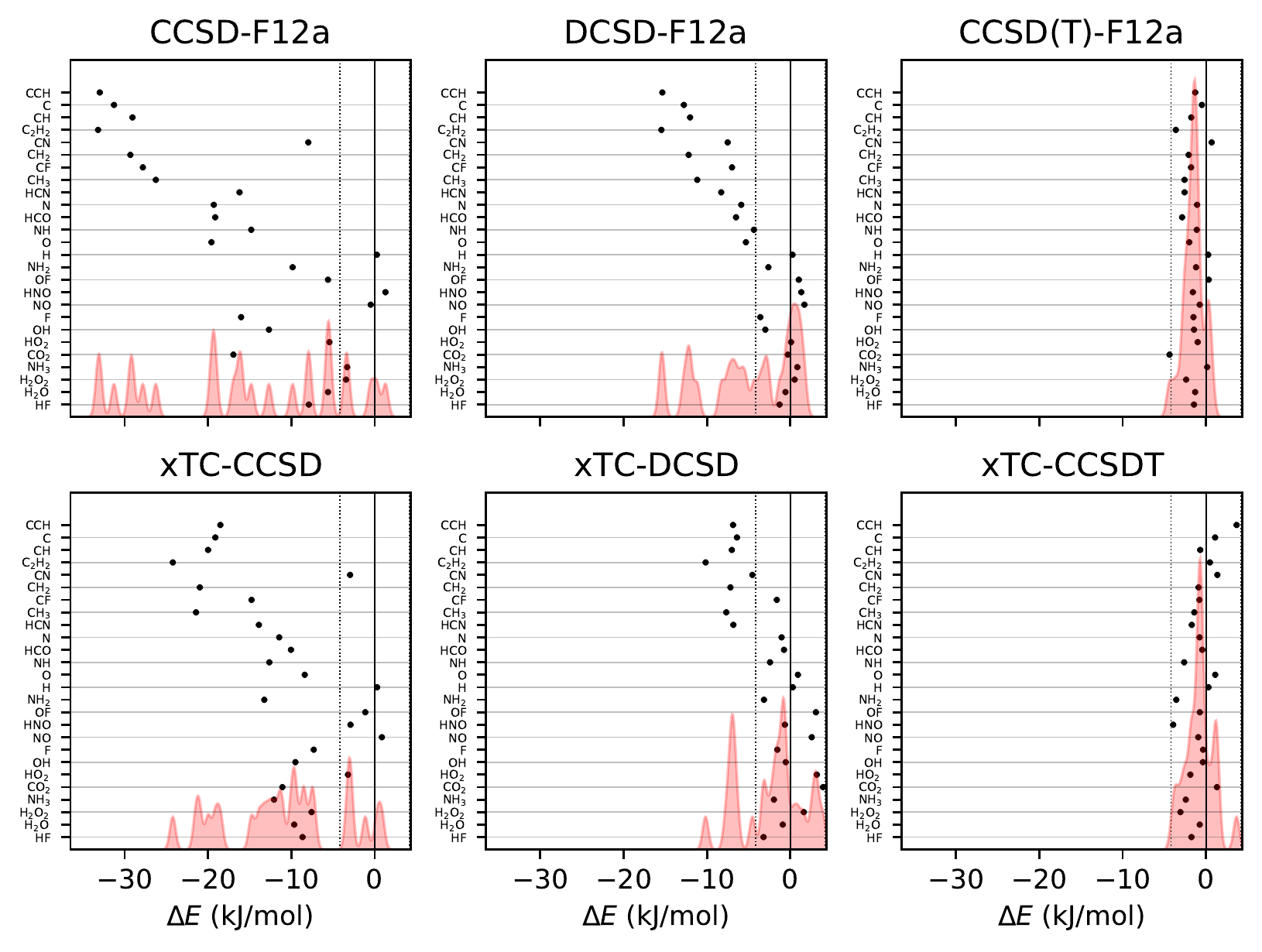}
\caption{\label{plt:methods:form} Method dependence of formation energies (aug-cc-pVTZ) compared to HEAT.
    Dotted lines indicate chemical accuracy.
    The shaded area corresponds to the sum of gaussians centered at the data points
    with the width of the gaussians chosen such that equidistantly distributed gaussians would be \chg{contained} to 95\% in the corresponding segment.
    }
\end{figure*}
\begin{table*}
\caption{\label{tab:methods}Method dependence of energies (aug-cc-pVTZ) compared to HEAT.}
\begin{ruledtabular}
\begin{tabular}{rrrrrrrrrr}
&\multicolumn{3}{c}{$E_{\rm tot}$ (mhartee)}&\multicolumn{3}{c}{$\Delta E_{\rm at}$ (kJ/mol)}&\multicolumn{3}{c}{$\Delta E_{\rm form}$ (kJ/mol)}\\
\cline{2-4}
\cline{5-7}
\cline{8-10}
&{ME}&{STD}&{MaxE}&{MAE}&{RMSE}&{MaxE}&{MAE}&{RMSE}&{MaxE}\\
\hline
{CCSD-F12a}&\chg{28.5}&\chg{16.3}&\chg{62.5}&\chg{23.91}&\chg{28.34}&\chg{-53.47}&\chg{15.21}&\chg{18.58}&\chg{-33.18}\\
{DCSD-F12a}&\chg{20.0}&\chg{10.9}&\chg{44.0}&\chg{7.82}&\chg{9.50}&\chg{-23.13}&\chg{5.42}&\chg{7.32}&\chg{-15.47}\\
{CCSD(T)-F12a}&\chg{16.4}&\chg{8.7}&\chg{32.9}&\chg{1.47}&\chg{1.90}&\chg{-4.01}&\chg{1.60}&\chg{1.89}&\chg{-4.37}\\
\chg{VMC}&\chg{105.3}&\chg{ 61.9}&\chg{206.1}&\chg{ 102.50}&\chg{ 126.81}&\chg{-221.50}&\chg{  88.67}&\chg{ 105.30}&\chg{-204.19}\\
\chg{xTC-HF}&\chg{181.2}&\chg{105.3}&\chg{373.5}&\chg{193.37}&\chg{230.16}&\chg{-374.34}&\chg{143.38}&\chg{166.77}&\chg{-285.83}\\
{xTC-CCSD}&{13.3}&{8.9}&{30.9}&{11.36}&{14.17}&{-27.62}&{10.99}&{12.87}&{-24.21}\\
{xTC-DCSD}&{9.4}&{6.6}&{25.7}&{2.69}&{3.35}&{-8.49}&{3.47}&{4.39}&{-10.16}\\
{xTC-CCSDT}&{8.2}&{6.0}&{24.7}&{2.27}&{2.58}&{5.88}&{1.48}&{1.82}&{-3.94}\\
\end{tabular}
\end{ruledtabular}
\end{table*}
The results for transcorrelated and explicitly correlated F12a methods are shown in Table~\ref{tab:methods}
and Figures~\ref{plt:methods:mol}, \ref{plt:methods:at} and \ref{plt:methods:form}. 
\chg{Additionally, the variational Monte-Carlo (VMC) energies obtained in the Jastrow optimization
as well as the mean-field energies obtained with the xTC integrals (xTC-HF) have been included in Table~\ref{tab:methods}.}
Note that we have used CCSDT instead of CCSD(T) for the transcorrelated calculations since
use of the xTC ansatz would require an iterative approach to the perturbative inclusion of the triples.
Furthermore, there are several possibilities for non-iterative (T) corrections in transcorrelated methods,
which will be investigated in a separate publication.

For total energies, transcorelated methods are on average by a factor of \chg{roughly two} closer to the HEAT reference than the respective explicitly correlated methods.
For relative energies however, the F12a methods benefit to a much larger degree from error compensation than the transcorrelated methods.
Nevertheless, \chg{xTC-CCSD and xTC-DCSD} on average still perform better than their explicitly correlated counterparts.
While CCSD is unable to provide satisfactory results both with the transcorralated and the explicitly correlated ansatz,
DCSD performs significantly better.
xTC-DCSD in particular yields average errors close to or even within chemical accuracy (1 kcal/mol = 4.2 kJ/mol = 1.6 mhartree) 
that are \chg{however still about twice as large as} those of CCSD(T)-F12a.
\chg{Nevertheless, it provides} a reasonable compromise between accuracy and computational cost. 
xTC-CCSDT further improves both absolute and relative energies compared to xTC-DCSD,
so the success of xTC-DCSD is unlikely to be due to fortuitous error compensation alone.
\chg{CCSD(T)-F12a emerges as the overall best method for the treatment of relative energies, 
with xTC-CCSDT yielding slightly worse results for atomization energies 
and almost identical results for formation energies.
However}, unlike CCSD(T)-F12a where explicit correlation in the triples part enters only implicitly via the doubles,
inclusion of explicit correlation via xTC for higher order excitations is trivial 
in xTC-CCSDT and subsequent non-perturbative methods in the coupled cluster hierarchy
and comes with no additional computational cost beyond generation of the xTC integrals.

\subsection{Cardinal number and diffuse functions}
\begin{table*}
\caption{\label{tab:basis}Basis set dependence of xTC-DCSD energies compared to HEAT.}
\begin{ruledtabular}
\begin{tabular}{rrrrrrrrrr}
&\multicolumn{3}{c}{$E_{\rm tot}$ (mhartee)}&\multicolumn{3}{c}{$\Delta E_{\rm at}$ (kJ/mol)}&\multicolumn{3}{c}{$\Delta E_{\rm form}$ (kJ/mol)}\\
\cline{2-4}
\cline{5-7}
\cline{8-10}
&{ME} &{STD} &{MaxE}&{MAE}&{RMSE}&{MaxE}&{MAE}&{RMSE}&{MaxE}\\
\hline
{cc-pVDZ}&{ 64.7}&{ 39.5}&{148.5}&{41.26}&{45.67}&{-74.59}&{20.20}&{23.63}&{-44.96}\\
{aug-cc-pVDZ}&{ 49.0}&{ 29.7}&{108.1}&{31.18}&{36.51}&{-69.50}&{18.69}&{24.15}&{-51.58}\\
{cc-pVTZ}&{13.7}&{ 9.1}&{35.4}&{5.59}&{6.39}&{-11.37}&{3.97}&{4.73}&{-9.36}\\
{aug-cc-pVTZ}&{ 9.4}&{ 6.6}&{25.7}&{2.69}&{3.35}&{-8.49}&{3.47}&{4.39}&{-10.16}\\
{cc-pVQZ}&{3.6}&{2.8}&{8.6}&{6.00}&{6.90}&{-13.14}&{4.13}&{5.31}&{-11.76}\\
{aug-cc-pVQZ}&{2.8}&{2.3}&{7.5}&{5.01}&{6.13}&{-13.40}&{3.91}&{5.31}&{-12.30}\\
\end{tabular}
\end{ruledtabular}
\end{table*}
A comparison of the results for xTC-DCSD with the basis sets cc-pVXZ and aug-cc-pVXZ (X = D, T, Q)\cite{Dunning1989} can be found in Table~\ref{tab:basis}.
The total energies become progressively better with increasing cardinal number, but on average do not reach mhartree accuracy.
Inclusion of diffuse functions further improves the results
for the total energies, but the average errors remain larger than for the unaugmented basis set with the next higher cardinal number.

This clear trend does not translate to the atomization and formation energies.
The double zeta basis sets yield results with average errors of up to roughly ten times chemical accuracy and are therefore clearly unsuitable.
Triple zeta basis sets on the other hand are already able to achieve chemical accuracy
for many molecules, especially if diffuse functions are included.
However, despite the further improvement in the total energies for quadruple zeta basis sets, no further improvement 
in the relative energies is observed. Instead, the results become slightly worse while still remaining close to chemical accuracy,
indicating that we benefit from favorable error compensation for the triple zeta basis sets.

\subsection{Core correlated basis sets}
\begin{table*}
\caption{\label{tab:core}Effect of core-correlated basis sets on xTC-DCSD energies compared to HEAT.}
\begin{ruledtabular}
\begin{tabular}{rrrrrrrrrr}
&\multicolumn{3}{c}{$E_{\rm tot}$ (mhartee)}&\multicolumn{3}{c}{$\Delta E_{\rm at}$ (kJ/mol)}&\multicolumn{3}{c}{$\Delta E_{\rm form}$ (kJ/mol)}\\
\cline{2-4}
\cline{5-7}
\cline{8-10}
&{ME} &{STD} &{MaxE}&{MAE} &{RMSE}&{MaxE}&{MAE} &{RMSE}&{MaxE}\\
\hline
{cc-pwCVDZ}&{ 55.1}&{ 36.2}&{135.9}&{ 40.41}&{ 45.42}&{-81.43}&{ 20.27}&{ 23.50}&{-42.29}\\
{aug-cc-pwCVDZ}&{39.7}&{26.1}&{95.9}&{ 30.64}&{ 35.71}&{-67.56}&{ 17.83}&{ 22.71}&{-47.84}\\
{cc-pVTZ}&{13.7}&{ 9.1}&{35.4}&{5.59}&{6.39}&{-11.37}&{3.97}&{4.73}&{-9.36}\\
{aug-cc-pVTZ}&{ 9.4}&{ 6.6}&{25.7}&{2.69}&{3.35}&{-8.49}&{3.47}&{4.39}&{-10.16}\\
{cc-pwCVTZ}&{11.5}&{ 7.7}&{29.2}&{  8.36}&{  9.39}&{-15.35}&{  4.49}&{  5.15}&{-10.37}\\
{aug-cc-pwCVTZ}&{ 7.7}&{ 5.4}&{20.3}&{  5.28}&{  7.24}&{-17.02}&{  4.40}&{  5.98}&{-13.06}\\
\end{tabular}
\end{ruledtabular}
\end{table*}
The results for the use of the core-correlated basis sets cc-pwCVXZ and aug-cc-pwCVXZ (X = D, T)\cite{Dunning2002} are shown in Table~\ref{tab:core}.
For double zeta basis sets, we see improvement in both total and relative energies, but not nearly enough
to justify their use. For triple zeta basis sets, the total energies improve compared to the respective basis
sets without additional core functions. For atomization energies however, use of core-correlated basis sets
leads to larger errors outside of chemical accuracy and aug-cc-pVTZ results remain closest to the reference
for the systems under investigation. For formation energies, the deterioration of results is less pronounced and the 
results remain close to chemical accuracy.

\subsection{Effect of F12 basis sets}
\begin{table*}
\caption{\label{tab:f12}Effect of F12 basis sets on xTC-DCSD energies compared to HEAT.}
\begin{ruledtabular}
\begin{tabular}{rrrrrrrrrr}
&\multicolumn{3}{c}{$E_{\rm tot}$ (mhartee)}&\multicolumn{3}{c}{$\Delta E_{\rm at}$ (kJ/mol)}&\multicolumn{3}{c}{$\Delta E_{\rm form}$ (kJ/mol)}\\
\cline{2-4}
\cline{5-7}
\cline{8-10}
&{ME} &{STD} &{MaxE}&{MAE} &{RMSE}&{MaxE}&{MAE} &{RMSE}&{MaxE}\\
\hline
{cc-pVDZ-F12}&{23.8}&{15.8}&{57.0}&{ 18.57}&{ 22.94}&{-50.61}&{ 11.02}&{ 14.85}&{-31.19}\\
{aug-cc-pVDZ-F12}&{23.2}&{15.5}&{55.8}&{ 17.99}&{ 22.41}&{-50.94}&{ 11.54}&{ 15.39}&{-31.55}\\
{cc-pVTZ}&{13.7}&{ 9.1}&{35.4}&{5.59}&{6.39}&{-11.37}&{3.97}&{4.73}&{-9.36}\\
{aug-cc-pVTZ}&{ 9.4}&{ 6.6}&{25.7}&{2.69}&{3.35}&{-8.49}&{3.47}&{4.39}&{-10.16}\\
{cc-pVTZ-F12}&{2.6}&{2.4}&{7.5}&{ 5.13}&{ 6.42}&{-14.36}&{  4.00}&{  5.56}&{-12.75}\\
{aug-cc-pVTZ-F12}&{2.5}&{2.4}&{7.6}&{ 4.71}&{ 6.00}&{-14.03}&{  3.83}&{  5.37}&{-12.48}\\
\end{tabular}
\end{ruledtabular}
\end{table*}
Results for the basis sets cc-pVXZ-F12 and aug-cc-pVXZ-F12 (X = D, T)\cite{Kirk:VnZF1208,aVXZF12} optimized for use with explicitly correlated methods are shown in Table~\ref{tab:f12}.
While total energies are greatly improved compared to the respective basis sets discussed previously,
the relative energies paint an ambiguous picture. Atomization and formation energies for double zeta basis sets
are greatly enhanced but still far from chemical accuracy. For triple zeta basis sets however we observe slight deterioration of the results
despite the more accurate total energies.
Therefore, we come to the conclusion that the triple zeta F12 basis sets offer no advantage here
compared to the smaller aug-cc-pVTZ basis set.

\subsection{Frozen core approximation}
\begin{table*}
\caption{\label{tab:frozencore}Effect of frozen core approximation on energies (aug-cc-pVTZ) compared to HEAT.}
\begin{ruledtabular}
\begin{tabular}{rrrrrrrrrr}
&\multicolumn{3}{c}{$E_{\rm tot}$ (mhartee)}&\multicolumn{3}{c}{$\Delta E_{\rm at}$ (kJ/mol)}&\multicolumn{3}{c}{$\Delta E_{\rm form}$ (kJ/mol)}\\
\cline{2-4}
\cline{5-7}
\cline{8-10}
&{ME} &{STD} &{MaxE}&{MAE}&{RMSE}&{MaxE}&{MAE}&{RMSE}&{MaxE}\\
\hline
{FC-xTC-DCSD}&{102.0}&{48.0}&{210.3}&{9.65}&{13.02}&{-26.18}&{6.27}&{8.19}&{-19.72}\\
{xTC-FC-DCSD}&{11.3}&{6.7}&{25.8}&{3.79}&{5.06}&{-12.13}&{3.68}&{4.80}&{-12.02}\\
{xTC-DCSD}&{9.4}&{6.6}&{25.7}&{2.69}&{3.35}&{-8.49}&{3.47}&{4.39}&{-10.16}\\
{FC-DCSD-F12a}&\chg{98.7}&\chg{46.9}&\chg{204.9}&\chg{12.79}&\chg{15.16}&\chg{-35.50}&\chg{6.09}&\chg{8.13}&\chg{-19.27}\\
{DCSD-F12a}&\chg{20.0}&\chg{10.9}&\chg{44.0}&\chg{7.82}&\chg{9.50}&\chg{-23.13}&\chg{5.42}&\chg{7.32}&\chg{-15.47}\\
{FC-CCSD(T)-F12a}&\chg{95.3}&\chg{44.4}&\chg{194.0}&\chg{5.47}&\chg{6.53}&\chg{-12.78}&\chg{1.91}&\chg{2.57}&\chg{-7.05}\\
{CCSD(T)-F12a}&\chg{16.4}&\chg{8.7}&\chg{32.9}&\chg{1.47}&\chg{1.90}&\chg{-4.01}&\chg{1.60}&\chg{1.89}&\chg{-4.37}\\
\end{tabular}
\end{ruledtabular}
\end{table*}
An obvious approach to reduce the cost of the calculation is the frozen core approximation.
The orbitals can be frozen either before generating the additional integrals (FC-xTC) 
or after calculating the xTC corrections (xTC-FC). 

In the first approach, we employ the standard frozen core approximation after evaluation of the one- and two-electron integrals
eqs. (\ref{eq:Inth}) and (\ref{eq:IntV}), respectively,  
and evaluate the transcorrelated integrals only for the remaining orbital space.
However, this means that we simply dismiss the contribution of the core orbitals to the transcorrelation ansatz.

For the xTC-FC approach on the other hand, we first evaluate the xTC corrections to the integrals
and only afterwards apply the frozen core approximation according to the standard equations,
leading to inclusion of additional correlation arising from the core orbitals.
Note that for xTC-FC, the one-electron integrals are no longer hermitian due to the contraction
and addition of non-hermitian two-electron integrals onto the one-electron integrals.

While FC-xTC obviously reduces the cost of the generation of the transcorrelated integrals,
the effect on the efficiency of the subsequent correlation treatment is identical for both approaches. 
Being able to employ the FC-xTC approach would enable treatment of larger systems,
but that would require the additional correlation included in xTC-FC via the transcorrelated integrals to be negligible.

The results for both approaches can be found in Table~\ref{tab:frozencore} together with results for DCSD-F12a and CCSD(T)-F12a with and without frozen core approximation. For all molecules,
the 1s-shell was frozen for each C, N, O and F atom in the respective molecule.
The FC-xTC approximation increases the average error in the total energies by an order of magnitude
and the average errors in the atomization and formation energies by a factor of about three and two, respectively.
For the F12a methods use of the frozen core approximation also significantly deteriorates the quality of the total energies.
The effect on the atomization \chg{and formation} energies however is far less \chg{severe,
showing} that the F12 approach benefits again from excellent error compensation.

The total energies for FC-xTC and the FC-F12a methods are remarkably close to each other (see also the additional data provided in the supplementary material)
despite the all electron calculations providing vastly different results.
The xTC-FC approach however introduces comparatively small errors in both total and relative energies.
This indicates that most of the correlation due to core-core and core-valence interactions 
is already included in the xTC ansatz and the subsequent correlation treatment primarily
accounts for the valence-valence interactions.

\subsection{Density and orbital choice}
\begin{table*}
\caption{\label{tab:orbitals}Effect of using spin-averaged (SA) approach for open-shell systems and choice of orbitals and density on xTC-DCSD energies (aug-cc-pVTZ) compared to HEAT.}
\begin{ruledtabular}
\begin{tabular}{rrrrrrrrrr}
&\multicolumn{3}{c}{$E_{\rm tot}$ (mhartee)}&\multicolumn{3}{c}{$\Delta E_{\rm at}$ (kJ/mol)}&\multicolumn{3}{c}{$\Delta E_{\rm form}$ (kJ/mol)}\\
\cline{2-4}
\cline{5-7}
\cline{8-10}
&{ME} &{STD} &{MaxE}&{MAE}&{RMSE}&{MaxE}&{MAE}&{RMSE}&{MaxE}\\
\hline
{HF/HF}&{ 9.4}&{ 6.6}&{25.7}&{2.69}&{3.35}&{-8.49}&{3.47}&{4.39}&{-10.16}\\
{HF/HF(SA)}&{ 9.6}&{ 6.5}&{25.7}&{2.67}&{3.09}&{6.14}&{2.52}&{3.02}&{6.14}\\
{HF/DCSD}&{ 9.2}&{ 6.4}&{25.1}&{2.23}&{2.70}&{-6.44}&{3.35}&{4.22}&{-8.42}\\
{HF/DCSD(SA)}&{ 9.4}&{ 6.3}&{25.1}&{3.28}&{3.99}&{8.69}&{2.38}&{2.90}&{7.36}\\
{DCSD/DCSD}&{ 9.4}&{ 6.5}&{25.4}&{2.73}&{3.53}&{-10.49}&{4.37}&{5.65}&{-12.13}\\
{DCSD/DCSD(SA)}&{ 9.7}&{ 6.4}&{25.4}&{2.93}&{3.53}&{8.40}&{3.03}&{3.82}&{8.59}\\
\end{tabular}
\end{ruledtabular}
\end{table*}
The xTC ansatz offers some additional flexibility through the choice of the density matrix used in the contractions.
We are going to investigate three cases: HF orbitals with HF density (HF/HF), HF orbitals with DCSD density (HF/DCSD)
and DCSD natural orbitals with DCSD density (DCSD/DCSD).
Furthermore, it would be appealing to avoid using the restricted open-shell algorithm (appendix~\ref{sec:ROHFwork})
for open-shell molecules, since it leads not only to increased computational cost compared to the restricted closed-shell algorithm (appendix~\ref{sec:RHFwork})
but also to spin-specific one- and two-body integrals,
requiring additional storage space and the subsequent correlation methods to be able to handle spin-specific integrals.
This can be accomplished by using 
the density
of an open-shell system as the density in the closed-shell algorithm,
which we will refer to as the spin-averaged (SA) approximation.

The results for the three orbital and density choices with and without SA approximation are listed in Table~\ref{tab:orbitals}.
The orbital and density choice has only a marginal impact on the total energies.
For atomization and formation energies, HF/DCSD and DCSD/DCSD yield slightly better or worse results, respectively, on average.
These changes are unlikely to be meaningful and are likely a result of changes in error cancellation.

Curiously, using the SA approximation does not significantly change the quality of either total or 
relative energies. It can even lead to slightly better results, as can be seen for the HF/HF orbital choice using the SA approximation,
which shows the overall best results for the choices investigated here. 
Therefore, considering that other error sources in the current approach lead to significantly larger deviations, there is no compelling practical reason 
not to use the SA approximation.

\section{\label{sec:conclusion}Conclusion and Outlook}
We have implemented an approximation to the full transcorrelation treatment which excludes the full three-body term
of the normal ordered transcorrelated Hamiltonian (xTC) and includes the effective zero-, one- and two-body
contributions of the three-body term via modification of the remaining integrals and the nuclear repulsion term.
Explicit calculation of three-body integrals is avoided by obtaining the correction to the integrals
via contractions of a density matrix with the three-index intermediates for the numerical integration
of the three-body integrals, reducing the scaling by two orders of magnitude while introducing only minor errors.

The scheme presented here has the added benefit of no longer requiring explicit contractions 
with the three-body integrals in the subsequent correlation treatment, thus removing the need
for costly and complicated modifications in conventional correlation methods in order to allow the use of the transcorrelated method.
In contrast to F12 methods, inclusion of explicit correlation via xTC in 
\chg{multireference methods or} 
methods using triples and higher excitations is therefore trivial
\chg{as long as one accomodates non-hermitian integrals.}

Furthermore, unlike the frozen core approximation for F12 methods,
freezing the core orbitals after generation of the xTC integrals (xTC-FC) 
allows partial inclusion of explicit correlation for the core orbitals.
This has been shown to lead to only minor deviations from the all electron calculations
allowing for computational savings in the subsequent correlation treatment without significant loss in accuracy.

Benchmark calculations have been carried out on the molecules of the HEAT set and it has been demonstrated
that chemical accuracy for atomization and formation energies is within reach for the comparatively cheap
xTC-DCSD method and the aug-cc-pVTZ basis set. \chg{For methods that do not include triple excitations}, 
results are \chg{generally} better than those obtained
with the respective explicitly correlated F12 method
\chg{while methods that include triples excitations yield similar results for xTC and F12}.
While improving total energies, the more expensive basis sets (aug-)cc-pVQZ, (aug-)cc-pwCVTZ and (aug-)cc-pVTZ-F12
fail to show improvement in the relative energies compared to results obtained with aug-cc-pVTZ.

The xTC ansatz in combination with the DCSD method enables the application of transcorrelation to systems of several hundred orbitals
at a reasonable computational cost. Further development should focus on additional efficiency improvements
and the development of more balanced Jastrow factors in order to further improve the quality of relative energies.
Since the DCSD method has proven to be less unstable for systems with strong static correlation than
conventional coupled cluster methods, the study of strongly correlated systems may be an interesting application as well.

\chg{
\section*{Supplementary Material}
Raw data and additional figures can be found in the supplementary material.}

\begin{acknowledgments}
Funded by the Deutsche Forschungsgemeinschaft (DFG, German Research Foundation) -- 455145945. Financial support from the Max-Planck Society is gratefully acknowledged.
P.L.R. and A.A. acknowledge support from the European Centre of
Excellence in Exascale Computing TREX, funded by the Horizon 2020
program of the European Union under grant no.\ 952165.
Any views and opinions expressed are those of the authors only and do
not necessarily reflect those of the European Union or the European
Research Executive Agency.
Neither the European Union nor the granting authority can be held
responsible for them.
\end{acknowledgments}

\section*{Data Availability Statement}
The data that support the findings of this study are available within the article and its supplementary material.

\section*{Author Declarations}
The authors have no conflicts to disclose.

\section*{References}
\bibliography{bib}

\appendix

\section{Restricted open-shell working equations}
\label{sec:ROHFwork}
Integrating equations (\ref{eq:tcmf:1el}), (\ref{eq:tcmf:0el}), (\ref{eq:tcmf:2el}) and (\ref{eq:tcmf:inter}) over spin 
and assuming use of the same spatial orbitals for $\alpha$- and $\beta$-spin, we can simplify the working equations.
In our current implementation we also assume 
\begin{subequations}
\begin{align}
    &[\rho_{p}^{q}]^\sigma = \rho_{p}^{q} &&\forall \sigma\in\alpha,\beta
    \\
    &[\vec V_{p}^{q}]^{\sigma\tau} = \vec V_{p}^{q} &&\forall \sigma,\tau \in \alpha,\beta,
\end{align}
\end{subequations}
i.e. in addition to using the same orbitals for $\alpha$- and $\beta$-spin we also use the same Jastrow factors.
While this removes much of the spin-dependency, we still obtain spin-specific sets of one- and two-electron integrals
due to the use of spin-specific density matrices.

\begin{align}
    [\Delta h_{p}^{q}]^\sigma = 
    -\frac{1}{2} \bigg(
    &\sum_\tau [\Delta U_{pr}^{qs}]^{\sigma\tau}[\gamma^{r}_{s}]^\tau
    \nonumber\\&
    -[\Delta U_{pr}^{sq}]^{\sigma\sigma}[\gamma^{r}_{s}]^\sigma
    \bigg)     
\end{align}

\begin{eqnarray}
    \bra{\Psi_0}L\ket{\Psi_0} = -\frac{1}{3} \sum_\sigma [\Delta h_{p}^{q}]^\sigma [\gamma^{p}_{q}]^\sigma
\end{eqnarray}

\begin{align}
    [\Delta U_{pr}^{qs}]^{\sigma\tau} =
    -\mathcal P^{(pq\sigma)}_{(rs\tau)}
    \big(
        &\rho_{p}^{q}(i)[A_{r}^{s}(i)]^{\tau}
        \nonumber\\&
        +\vec V_{p}^{q}(i)\cdot\vec [B_{r}^{s}(i)]^{\tau}
    \big)
\end{align}
\begin{subequations}
\begin{eqnarray}
    \rho_{p}^{q}(i) =w_i\phi_p^*(i)\phi^q(i)
\end{eqnarray}
\begin{eqnarray}
    [A_{r}^{s}(i)]^{\tau} = \tilde V_{r}^{s}(i) - [\tilde Z_{r}^{s}(i)]^{\tau}
\end{eqnarray}
\begin{eqnarray}
    \tilde V_{r}^{s}(i) = \vec W(i)\cdot\vec V_{r}^{s}(i)
\end{eqnarray}
\begin{eqnarray}
    \vec W(i) = \vec V_{t}^{u}(i)\sum_\rho[\gamma_{u}^{t}]^\rho
\end{eqnarray}
\begin{eqnarray}
    [\tilde Z_{r}^{s}(i)]^{\tau} = \vec V_{r}^{u}(i)\cdot[\vec X_{u}^{s}(i)]^{\tau}
\end{eqnarray}
\begin{eqnarray}
    [\vec X_{u}^{s}(i)]^{\tau} = \vec V_{t}^{s}(i) [\gamma^{t}_{u}]^\tau
\end{eqnarray}
\begin{eqnarray}
    [\vec B_{r}^{s}(i)]^{\tau} = 
    \frac{1}{2}\tilde W(i)\vec V_{r}^{s}(i) 
    - [\vec Z_{r}^{s}(i)]^{\tau}
\end{eqnarray}
\begin{eqnarray}
    \tilde W(i) = \rho_{t}^{u}(i)\sum_\rho [\gamma^{t}_{u}]^\rho
\end{eqnarray}
\begin{eqnarray}
    [\vec Z_{r}^{s}(i)]^{\tau} = 
    \rho_{r}^{u}(i)[\vec X_{u}^{s}(i)]^{\tau} 
    + [\vec Y_{r}^{t}(i)]^{\tau} \rho_{t}^{s}(i)
\end{eqnarray}
\begin{eqnarray}
    [\vec Y_{r}^{t}]^{\tau}(i) = \vec V_{r}^{u}(i) [\gamma^{t}_{u}]^\tau
\end{eqnarray}
\end{subequations}

\section{Restricted closed-shell working equations}
\label{sec:RHFwork}
Further assuming a closed-shell system, we can drop the remaining spin-dependencies.
The following equations have been used for the closed-shell systems investigated in this study. 
They can double as a spin-averaged approximation to the restricted open-shell case
if the density matrices are taken as the average of $\alpha$- and $\beta$-density matrices. 
This can be desirable, since unlike the former equations, there are no spin-specific sets of integrals.

Further note that the equations given here deviate from the standard convention of including the factor of two from the spin-integration
into the density matrix. Instead this factor is explicitly included in the equations presented here. 

\begin{eqnarray}
    \Delta h_{p}^{q} = 
    -\frac{1}{2} \bigg(
    2\Delta U_{pr}^{qs}\gamma^{r}_{s}
    -\Delta U_{pr}^{sq}\gamma^{r}_{s}
    \bigg) 
\end{eqnarray}

\begin{eqnarray}
    \bra{\Psi_0}L\ket{\Psi_0} = -\frac{2}{3} \Delta h_{p}^{q} \gamma^{p}_{q}
\end{eqnarray}

\begin{align}
    \Delta U_{pr}^{qs} =
    -\mathcal P^{(pq)}_{(rs)}
    \big(
        &\rho_{p}^{q}(i)A_{r}^{s}(i)
        +
        \vec V_{p}^{q}(i)\cdot\vec B_{r}^{s}(i)
    \big)
\end{align}

\begin{subequations}
\begin{eqnarray}
    \rho_{p}^{q}(i) =w_i\phi_p^*(i)\phi^q(i)
\end{eqnarray}
\begin{eqnarray}
    A_{r}^{s}(i) = \tilde V_{r}^{s}(i) - \tilde Z_{r}^{s}(i)
\end{eqnarray}
\begin{eqnarray}
    \tilde V_{r}^{s}(i) = \vec W(i)\cdot\vec V_{r}^{s}(i)
\end{eqnarray}
\begin{eqnarray}
    \vec W(i) = 2\vec V_{t}^{u}(i)\gamma^{t}_{u}
\end{eqnarray}
\begin{eqnarray}
    \tilde Z_{r}^{s}(i) = \vec V_{r}^{u}(i)\cdot\vec X_{u}^{s}(i)
\end{eqnarray}
\begin{eqnarray}
    \vec X_{u}^{s}(i) = \vec V_{t}^{s}(i) \gamma^{t}_{u}
\end{eqnarray}
\begin{eqnarray}
    \vec B_{r}^{s}(i) = \frac{1}{2}\tilde W(i)\vec V_{r}^{s}(i) - \vec Z_{r}^{s}(i)
\end{eqnarray}
\begin{eqnarray}
    \tilde W(i) = 2\rho_{t}^{u}(i)\gamma^{t}_{u}
\end{eqnarray}
\begin{eqnarray}
    \vec Z_{r}^{s}(i) = \rho_{r}^{u}(i)\vec X_{u}^{s}(i) + \vec Y_{r}^{t}(i) \rho_{t}^{s}(i)
\end{eqnarray}
\begin{eqnarray}
    \vec Y_{r}^{t}(i) = \vec V_{r}^{u}(i) \gamma^{t}_{u}
\end{eqnarray}
\end{subequations}

\end{document}